\documentstyle[11pt]{article}
\input{psfig}
\newtheorem{theorem}{\bf Theorem}
\newtheorem{lemma}{\bf Lemma}
\newtheorem{corollary}{\bf Corollary}

\newtheorem{definition}{\bf Definition}
\newtheorem{claim}{\bf Claim}

\newenvironment{proof}{\par \bf Proof. \rm}{$\Box$ \vspace{1ex}}

\date{}

\title{Reversibility and Adiabatic
Computation: Trading Time and Space for Energy\thanks{Parts
of this paper were presented in preliminary form
in {\em Proc. IEEE Physics 
of Computation Workshop},
Dallas (Texas), Oct. 4-6, 1992, pp. 42-46, and
{\em  Proc. 11th IEEE Conference on Computational Complexity},
Philadelphia (Pennsylvania),  May 24-27,  1996.}}
\author{
Ming Li\thanks{Supported in part by
NSERC operating grant OGP-046506, ITRC, and a CGAT grant.
Address: Computer Science Department,
University of Waterloo,
Waterloo, Ontario, Canada N2L 3G1. Email: mli@math.uwaterloo.ca}\\
University of Waterloo\\
\and
Paul Vit\'{a}nyi\thanks{Partially
supported by the European Union
through NeuroCOLT ESPRIT Working Group Nr. 8556,
and by  NWO through NFI Project ALADDIN under Contract
number NF 62-376 and
NSERC under
International Scientific Exchange Award ISE0125663.
Address: CWI,
Kruislaan 413, 1098 SJ Amsterdam, The Netherlands. Email: paulv@cwi.nl}\\
CWI and University of Amsterdam\\}

\begin{document}
\maketitle
\begin{abstract}
Future miniaturization  and mobilization 
of computing devices requires
energy parsimonious
`adiabatic' computation. This
is contingent on
logical reversibility of computation.
An example is the idea of quantum computations
which are reversible except for the irreversible
observation steps.
We propose to
study quantitatively the exchange of 
computational resources like time and space for
irreversibility in computations.
Reversible simulations
of irreversible computations are memory intensive.
Such (polynomial time) simulations are analysed
here in terms of `reversible' pebble games. 
We show that Bennett's pebbling strategy uses least additional space
for the greatest number of simulated steps.
We derive a trade-off for storage space versus irreversible erasure.
Next we consider reversible computation itself.
An alternative
proof is provided for the precise expression
of the ultimate irreversibility
cost of an otherwise reversible computation without
restrictions on time and space use.
A time-irreversibility trade-off hierarchy 
in the exponential time region is exhibited.
Finally, extreme
time-irreversibility trade-offs for reversible
computations in the thoroughly unrealistic
range of computable versus noncomputable time-bounds are given.
\end{abstract}

\section{Introduction}
The ultimate limits of miniaturization of computing
devices, and therefore the speed of computation,
are constrained
by  the increasing
density of switching elements in the device. 
Linear speed up
by shortening interconnects on a two-dimensional device
is attended by cubing
the dissipated energy per area unit per second. Namely, we square
the number of switching elements
per area unit and linearly increase the number of switching
events per switch per time unit. The attending energy dissipation
on this scale in the long run cannot be compensated for by
cooling.
Reduction of the energy dissipation
per elementary computation step therefore determines future advances
in computing power.  In view of
the difficulty in improving low-weight small-size battery
performance, low-energy computing is already at this time of writing a 
main determining factor in
advanced mobilization
of computing and communication. 

Since 1940 the
dissipated energy per bit operation in a computing
device has with remarkable
regularity decreased
by roughly one order of magnitude (tenfold) every five years, \cite{Ke88,La88}.
Extrapolations of current trends
show that the energy dissipation per
binary logic operation needs to be reduced below $kT$
(thermal noise)
within 20 years. Here $k$ is Boltzmann's constant and $T$
the absolute temperature in degrees Kelvin,
so that $kT \approx 3 \times 10^{-21}$
Joule at room temperature. Even at $kT$ level, 
a future device containing $10^{18}$ gates in a cubic centimeter operating at
a gigahertz dissipates about 3 million watts/second. For thermodynamic reasons,
cooling the operating temperature of such a computing device
to almost absolute zero (to get $kT$ down) must
dissipate at least as much energy in the cooling as it saves
for the computing, \cite{Me93}.

Considerations of thermodynamics of
computing started 
in the early fifties. J. von Neumann reputedly thought
that a computer operating at temperature $T$ must
dissipate at least $k T \ln 2$ Joule per elementary bit 
operation,
\cite{Bu}.
But R. Landauer~\cite{La61}
demonstrated
that it is only the `logically 
irreversible' operations in a physical computer
that are required to dissipate energy by generating a
corresponding amount of entropy for each bit of information
that gets irreversibly erased. As a consequence,
any arbitrarily
large reversible computation can be performed on an appropriate
physical device using only
one unit of physical energy {\em in principle}.  

Examples of logically reversible operations
are `copying' of records,
and `canceling' of one record with respect to
an identical record provided it is known that they
are identical. They are physically realizable (or almost
realizable) without energy dissipation.
Such operations occur when a program sets $y := x$
and later (reversibly) erases $x := 0$ while retaining
the same value in $y$. We shall
call such reversible erasure `canceling' $x$ against $y$. 
Irrespective of the original
contents of variable $x$ we can always restore $x$
by $x := y$. However, if the program has no copy of the value in
variable $x$
which can be identified by examining the
program without knowing the contents of
the variables, then after (irreversibly) erasing $x := 0$ 
we cannot restore the original contents
of $x$ even
though some variable $z$ may have by chance the same contents.
`Copying' and `canceling' are logically reversible, and
their energy dissipation free execution gives substance
to the idea that logically reversible computations
can be performed with zero energy dissipation.

Generally, an operation is {\it logically reversible}
if its inputs can always be deduced from the outputs.
Erasure of information in a way such that it cannot be retrieved
is not reversible. Erasing a bit irreversibly
necessarily dissipates  $k T \ln 2$ energy
in a computer operating at temperature $T$.
In contrast, computing in a logically reversible way
says nothing about whether or not the computation
dissipates energy.
It merely means that the laws of physics do not
require such a computer to dissipate energy.
Logically reversible computers built
from reversible circuits, \cite{FT82}, or the reversible
Turing machine, \cite{Be82}, implemented
with current technology will presumably dissipate energy 
but may conceivably be implemented by future technology 
in an adiabatic fashion.
Current conventional electronic technologies
for implementing `adiabatic' logically reversible computation 
are discussed in {\sloppy \cite{Me93,PC}}. 

An example of a hypothetical reversible computer 
that is both logically and physically 
perfectly reversible
and perfectly free from energy dissipation is the billiard ball
computer, \cite{FT82}. Another example is the exciting prospect of
quantum computation, \cite{Fe85,De85,Sh94},
which is reversible except for the irreversible
observation steps. 

\subsection{Outline of the Paper}
Here we propose the quantitative study of
exchanges of computing resources such as time and space
for irreversibility which we believe will be
relevant for the physics of future computation devices.

{\bf Reversible simulation.}
Bennett \cite{Be89} gives a general reversible simulation for
irreversible algorithms
in the stylized form of a pebble game.
While such
reversible simulations incur little overhead in additional
computation time, they may use a large amount
of additional memory space during the computation.
We show that among all simulations which can be modelled by
the pebble game, Bennett's simulation is optimal in that it
uses the least auxilliary space for the greatest number of
simulated steps. That is,
if $S$ is the space used by the simulated
irreversible computation, then the simulator
uses $nS$ space to simulate $(2^n-1)S$ steps of the simulated
computation.
Moreover, we show that no simple generalization of
such simulations can simulate
that many steps using $(n-1)S$ space.
On the other hand, we show that at the cost of a limited amount
of erasure the simulation can be made more space efficient:
we can save $kS$ space
in the reversible simulation at a cost of $(2^{k+2}-1)S$
irreversible bit erasures, for all $k$ with $1 \leq k \leq n$.
Hence there can be an
advantage in adding limited irreversibility
to an otherwise reversible simulation of conventional
irreversible computations. This may be
of some practical relevance for adiabatic computing.

{\bf Reversible computation.}
Next, we consider irreversibility issues
related to reversible computations themselves.
Such computations may be directly programmed on a reversible
computer or may be a reversible simulation of an irreversible
computation.
References \cite{Le63,Be73} show independently
that all computations can be performed logically reversibly 
at the cost of eventually filling up the memory with
unwanted garbage information. This means that
reversible computers with bounded memories
require in the long run irreversible bit operations,
for example, to erase records irreversibly to 
create free memory space.
The minimal possible number of irreversibly 
erased bits to do so determines
the ultimate limit of heat dissipation 
of the computation by Landauer's principle.

To establish the yardstick for subsequent trade-offs, 
we give an alternative direct operational proof for the known
exact expression of the ultimate number of irreversible bit
operations in an otherwise reversible computation,
without any bounds on computational resources like
time and space,
Theorem~\ref{thermo.distance}.\footnote{This is 
the unpublished proof 
in \cite{LV}; compare with the proof in \cite{BGLVZ93}.}

{\bf Time-Irreversibility trade-offs.}
Clearly, to potentially reduce physical energy dissipation
one first  needs to  reduce the number of irreversible
bit erasures in an otherwise reversible computation.
This can be achieved by
using more computation steps to drive
the number of irreversible computation steps closer to
ultimate limits. The method typically 
reversibly compresses `garbage' information before irreversibly
erasing it. (A similar situation holds
for space bounds on memory use.) 

{\bf Time-Irreversibility hierarchy.}
For exponential time bounds 
diagonalization techniques are used
to establish the existence of a sequence of increasing time bounds
for a computation resulting in a sequence of decreasing
irreversibility costs. (These time bounds are
exponential functions, while practical adiabatic computation
usually deals with less-than-exponential time in the size
of the input.)

{\bf Extreme trade-offs.}
In the thoroughly unrealistic
realm of computable versus noncomputable time-bounds it turns
out that there exist most  extreme
time-irreversibility trade-offs.

\subsection{Previous Work}
Currently, we are used to design
computational procedures containing irreversible operations.
To perform the intended computations
 without energy dissipation the related computation procedures need to become
completely reversible.
Fortunately, all irreversible computations 
can be simulated in a reversible manner, \cite{Le63,Be73}.
All known reversible  simulations of 
irreversible computations use little overhead in time
but large amounts of additional space. 
Commonly, polynomial time computations are considered as
the practically relevant
ones. Reversible simulation will not change such a time bound
significantly, but requires 
considerable additional memory space.
In this type of simulation one
needs to save on space; time is already almost optimal.

The reversible simulation in \cite{Be73} of $T$ steps of an
irreversible computation from $x$ to $f(x)$
reversibly  computes from input $x$ 
to output $\langle x, f(x) \rangle$
in $T' = O(T)$ time.
However, since this reversible simulation at some time instant
has to record the entire
history of the irreversible computation, its space use increases
linear with the number of simulated steps $T$. That is,
if the simulated irreversible computation uses $S$ space, then
for some constant $c > 1$ the simulation uses 
$T'\approx c+cT$ time and $S'\approx c + c(S+T)$ space.
The question arises whether one can reduce
the amount of auxiliary space needed by the simulation by a
more clever simulation method or
by allowing limited amounts of irreversibility.

In \cite{Be89} another elegant simulation technique is devised
reducing the auxiliary storage space.
This simulation does not save the entire history of the irreversible
computation but it breaks up the simulated computation
into segments of about $S$ steps
 and saves in a hierarchical manner {\em checkpoints}
consisting of complete instantaneous descriptions of the
simulated machine (entire tape contents, tape heads positions,
state of the finite control). After a later checkpoint is
reached and saved, the simulating machine reversibly
undoes its intermediate computation reversibly erasing
the intermediate history and reversibly canceling the previously
saved checkpoint. Subsequently, the computation is resumed from
the new checkpoint onwards. 

The reversible computation simulates $k^n$ segments of length $m$
of irreversible
computation in $(2k-1)^n$ segments of length $\Theta (m+S)$
of reversible computation using
$n(k-1)+1$ checkpoint registers using $\Theta (m+S)$
space each, for each $k,n,m$.
 
This way it is established that there are various trade-offs
possible in time-space in between $T'= \Theta (T)$ and
$S' = \Theta (TS)$ at one extreme ($k=1, m=T, n=1$) and (with the corrections
of \cite{LeSh90})
$T' = \Theta (T^{1+\epsilon}/S^{\epsilon} )$
and $S'= \Theta ( c(\epsilon) S(1+ \log T/S))$
with $c(\epsilon)= \epsilon 2^{1/\epsilon}$
for each $\epsilon > 0$,
using always the same simulation method but with different
parameters $k,n$ where $\epsilon = \log_k (2k-1)$ and $m = \Theta (S)$.
Typically, for $k=2$ we have $\epsilon = \log 3$.
Since for $T > 2^S$ the machine goes into
a computational loop, we always have $S \leq \log T$.
Therefore, it follows from Bennett's simulation
that  each irreversible Turing machine
using space $S$ can be simulated by a reversible machine
using space $S^2$ in polynomial time.

\section{Reversible Simulation}
Analysing the
simulation method of \cite{Be89} 
shows that it is essentially no better than
the simple \cite{Be73} simulation in terms of time
versus irreversible erasure trade-off. 
Extra irreversible erasing 
can reduce the simulation time of the former method to $\Theta (T)$,
but the `simple' method has $\Theta (T)$ simulation
time without irreversible
erasures anyway, but at the cost of a large space consumption. 
Therefore, it is crucial 
to decrease the extra space
required for the pure reversible simulation without 
increasing time if possible,
and in any case further reduce the extra space 
at the cost of limited numbers of irreversible erasures.

Since there is no better general reversible simulation
of an irreversible computation known as the above one,
and it seems likely that each proposed method must
have similar history preserving features,
analysis of this particular style of simulation may
in fact give results with more general validity.
We establish lower bounds on space use and upper bounds
on space versus irreversible erasure trade-offs.

To analyse such trade-offs we use Bennett's
brief suggestion in \cite{Be89}
that a reversible simulation can be modelled by the following
`reversible'  pebble game. Let $G$ be a linear list of
nodes $\{ 1,2, \ldots , T_G \}$.
We define a {\em pebble game} on $G$ as follows. The game
proceeds in a discrete sequence of steps of a single {\em player}.
There
are $n$ pebbles which can be put on nodes of $G$.
At any time the set of pebbles is divided in 
pebbles on nodes of $G$ and the remaining pebbles which are called
{\em free} pebbles. At each step either an existing
 free pebble can be put
on a node of $G$ (and is thus removed from the free pebble pool)
 or be removed from a node of $G$ (and is added to the
free pebble pool). The rules of the game are as follows.

\begin{enumerate}
\item
Initially $G$ is unpebbled and there is a pool of free pebbles.
\item
In each step the player can put either

(a) a free pebble on node $1$ or remove
a pebble from node $1$, or

(b) for some node $i>1$, put a free pebble on
node $i$ or remove a pebble from node $i$, provided
node $i-1$ is pebbled at the time.
\item
The player wins the game if he pebbles node $T_G$ and subsequently
removes all pebbles from $G$.
\end{enumerate}

The maximum number $n$ of pebbles 
which are simultaneously on $G$ 
at any time in the game gives the space complexity
$nS$ of the simulation. If one deletes a pebble not following
the above rules, then this means a block of bits of size $S$ is 
erased irreversibly. The limitation to
Bennett's simulation is in fact space, rather than time.
When space is limited, we may not have enough place to store garbage,
and these garbage bits will have to be irreversibly erased.
We establish a tight lower bound for {\em any}
strategy for the pebble game in order to obtain
a space-irreversibility trade-off.

\begin{lemma}\label{lem.pebble}
There is no winning strategy with $n$ pebbles for $T_G\geq 2^n$.
\end{lemma}

\begin{proof}
Fix any pebbling strategy for the player.
To prove the lemma it suffices to show that
the player
cannot reach node $f(k)=2^k$ using $k$ pebbles,
and also
remove all the pebbles at the end,
for $k:=1,2, \ldots$.
We proceed by induction.

{\em Basis: $k=1$.} It is straightforward to establish $f(1)=2$
cannot be reached with 1 pebble.

{\em Induction: $k \rightarrow k+1$.}
Assume that $f(i) = 2^i$ cannot be reached with $i$ pebbles, 
for $i=1, \ldots,k$, has been
established.
Consider
pebbling $G$ using $k+1$ pebbles.
Assume, that the player
can pebble node $f(k)+1=2^k+1$ (otherwise the induction
is finished).

Then, by the rules of the game,
there must be a {\it least} step $t$ such that for {\em all times}
$t'>t$ there are
pebbles on some nodes in
$f(k)+1, f(k)+2 , \ldots , T_G$. Among other things, this
implies that at step $t+1$ node
$f(k)+1$ is pebbled.

Partition the first
$f(k)-2$ nodes of $G$ into disjoint consecutive regions:
starting with node 1, region $L_i$ consists of 
the next block of  $f(k-i)$ nodes, for $i=1, \ldots, k-1$.
That is, $L_i = \{\sum_{j=1}^{k-i+1} 2^{k-j} +1, 
\ldots , \sum_{j=1}^{k-i} 2^{k-j} \}$. The
regions $L_1, \ldots , L_{k-1}$ cover nodes $1, \ldots , f(k)-2$.
Denote the remainder of $G$ but for nodes
$f(k)-1,f(k)$ by $R$, that is
 $R= G- \{f(k)-1, f(k)\}- \bigcup_{i=1}^{k-1} L_i = 
\{ f(k)+1, f(k)+2, \ldots , T_G \}$.

Consider the game from step $t+1$ onwards.
If there is always at least one pebble
on nodes $1, \ldots , f(k)$, then by inductive assumption 
the player can pebble with one initial pebble on $f(k)+1$
and the remaining $k-1$
free pebbles at most
$f(k)-1$ nodes and hence no further than
node $2f(k)-1 = 2^{k+1}-1$, and the induction is finished.

Therefore, 
to possibly pebble node $2^{k+1}$ the player
needs to remove all pebbles from nodes
$1, \ldots , f(k)$ first. Because node $f(k)+1$
was pebbled at step $t+1$, we know
that node $f(k)$ did have a pebble at that
time according to the game rules.
By assumption, from time $t+1$
there will henceforth always be
a leading pebble in region $R$.
Moreover, at time $t+1$ there is a pebble on node $f(k)$.
To remove all the pebbles in range $1, \ldots ,f(k)$,
the following requirements have to be satisfied.

\begin{itemize}
\item
From time $t+1$ onwards, there must always be a pebble at a
strategic location in $L_1$ until the
last remaining pebble in 
$G-(L_1 \cup R)= \{f(k-1)+1, \ldots , f(k)\}$ is removed.
Otherwise with at most 
$k-1$ pebbles, the player cannot
cross the unpebbled region $L_1$ (because
$|L_1|=f(k-1)$) to reach and
remove the {\em finally last remaining} pebble
in the range $G-(L_1 \cup R)$.
There are only $k-1$ pebbles available
because from time $t+1$ on we have a pebble in region $R$, 
and at least one pebble in $H= G - (L_1 \cup 
R)$.
\item
From time $t+1$ onwards, there must always be a pebble at a
strategic location in $L_2$ until the
last remaining pebble in $G-(L_1 \cup L_2 \cup R)=
   \{f(k-1) + f(k-2)+1, \ldots , f(k)\}$ is removed.
Otherwise, with at most 
$k-2$ pebbles, the player cannot
cross the unpebbled region $L_2$ (because
$|L_2|=f(k-2)$) to reach and
remove the {\em finally last remaining} 
pebble in the range $G-(L_1 \cup L_2 \cup R)$.
There are only $k-2$ pebbles available
because from time $t+1$ on we have a pebble in region $R$, a pebble
in $L_1$ (to help removing
the last remaining pebble in $L_2$), 
and at least one pebble in $H= G - (L_1 \cup L_2 \cup
R)$.
\item
By iteration of the argument,
there must be a pebble in each region $L_i$ at time $t+1$, 
for $i=1, \ldots , k-1$.
\end{itemize}

But these requirements use up $k-1$ pebbles located
in regions $L_1 , \ldots , L_{k-1}$. None of these regions
can become pebble-free before
we free the pebble on node $f(k)$, that is, the $k$th pebble. 
The $(k+1)$st pebble is in region $R$ forever after step $t+1$.
Therefore, there is no pebble left to pebble node $f(k)-1$ 
which is not
in $R \bigcup \{f(k)\} \bigcup_{i=1}^{k-1} L_i$. 
Hence it is impossible to remove all $k$ pebbles
from the first nodes $1, \dots ,f(k)$. Thus, leaving one pebble
in region $\{1, \ldots , f(k) \}$ with at most $k$ remaining pebbles,
by inductive assumption,
the player can pebble no farther than node $2f(k)-1$,
which 
finishes the induction.
\end{proof}

\begin{lemma}\label{lem.bennett}
There is a winning strategy with $n$ pebbles for $T_G= 2^n-1$.
\end{lemma}
\begin{proof}
Bennett's simulation \cite{Be89} is a winning strategy.
We describe his game strategy
as the pebble game $G= \{ 1, \ldots , T_G \}$,
recursively.
Let $I_k = I_{k-1} i_{k-1} I_{k-2} i_{k-2} \ldots I_1 i_1 I_0 i_0  $ 
where $I_j$ is
a sequence of $2^{j}-1$ consecutive locations in $G$,
and $i_j$ is the node directly following $I_j$,
for $j=0,1, \ldots ,k-1$. Note that $|I_0|=0$.

Let $F(k,I_k)$ be the program to pebble
an initially pebble-free interval $I_k$ of length $2^k-1$ of $G$, 
starting with $k$ free pebbles and a pebble-free 
$I_k$ and ending with $k$ pebbles on $I_k$ including
one pebble on the last node of $I_k$.

Let $F^{-1}(k,I_k)$ be the program starting with the end
configuration of $F(k,I_k)$ and executing the operation sequence
of $F(k,I_k)$ in reverse, each operation replaced by its
inverse which undoes what the original operation did,
ending with $F(k,I_k)$'s initial configuration.
We give the precise procedure in self-explanatory pseudo PASCAL. \\

\noindent
\begin{tabbing} 
{\bf Procedure}   $F(k,I_k)$: \\ 
 {\bf for} \= $i:=1,2, \ldots ,k$: \\
  \> $F(k-i, I_{k-i} )$; \\
 \>put pebble on node  $i_{k-i}$ ; \\
\> $F^{-1}(k-i, I_{k-i} )$ \\
\end{tabbing}

\noindent
\begin{tabbing}
{\bf Procedure}  $F^{-1}(k,I_k)$: \\ 
 {\bf for} \= $i:=k,k-1, \ldots ,1$: \\
 \> $F^{-1}(k-i, I_{k-i} )$; \\
 \>remove pebble on node  $i_{k-i}$ ; \\
 \>$F(k-i, I_{k-i} )$ \\
\end{tabbing}

Note that this way both $F(0, I_0)$ and $F^{-1}(0,I_0)$
are `skip' operations which don't change anything.
The size $T_G$ of a pebble game
which is won using this strategy using $n$ pebbles
is $|I_n|=2^n-1$. 
Moreover, if $F(k,I_k)$ takes $t(k)$ steps
we find $t(k)=2t(k-1) + \cdots +f(1)+k-1$. Then,
$t(k)=3 t(k-1) - 1$. That is, the number
of steps $T_G'$ of a winning play of a pebble game
of size $T_G=2^n-1$ is $T_G' \approx 3^n$, that is,
$T_G' \approx T_G^{\log 3}$.
\end{proof}

The simulation given in \cite{Be89} follows the rules
of the pebble game of length $T_G = 2^n-1$ with $n$ pebbles above.
A winning
strategy for a game of length $T_G$ using $n$ pebbles
corresponds with reversibly simulating $T_G$ segments of $S$
steps of an irreversible computation using $S$
space such that the reversible simulator
uses $T' \approx ST'_G \approx ST_G^{\log 3}$ steps 
and total space $S'=nS$. The space $S'$ corresponds
to the maximal number of pebbles on $G$
at any time during the game.  The placement or removal of a
pebble in the game corresponds to the reversible
copying or reversible cancelation of a `checkpoint'
consisting of the entire instantaneous description of size $S$
(work tape contents, location of heads, state of finite
control) of the simulated irreversible machine.
The total time $T_GS$ used by the irreversible computation
is broken up in segments of size $S$ so that the reversible
copying and canceling of a checkpoints takes about the same
number of steps as the computation segments in between
checkpoints.
\footnote{In addition to the rules of the pebble game 
there is a permanently pebbled initial node
so that the simulation actually uses $n+1$
pebbles for a pebble game with $n$ pebbles of length $T_G+1$.
The simulation uses $n+1 =S'/S$ pebbles for
a simulated number of $S(T_G+1)$ steps of the irreversible
computation.}

We can now formulate a trade-off between space used
by a polynomial time reversible computation and irreversible
erasures. First we show that allowing a limited
amount of erasure in an otherwise
reversible computation means that
we can get by with less work space.
Therefore, we define an {\em $m$-erasure} pebble game as
the pebble game above but with the additional rule

\begin{itemize}
\item
In at most $m$ steps
the player can 
remove a pebble from any node $i > 1$ without
node $i-1$ being pebbled at the time.
\end{itemize}
 
An $m$-erasure pebble game corresponds with an otherwise
reversible computation using $mS$ irreversible bit erasures,
where $S$ is the space used by the irreversible computation
being simulated.

\begin{lemma}\label{lem.erasure}
There is a winning strategy with $n$ pebbles
and $2m-1$ erasures for pebble games $G$
with $T_G= m2^{n-1}$, for all $m \geq 1$.
\end{lemma}
\begin{proof}
The strategy is to advance in blocks of size $2^{n-1}-1$
using $n-1$ pebbles without
erasures (as in Lemma~\ref{lem.bennett}),
put the $n$th pebble in front, and invert 
the advancement process to free all the
pebbles in the block. The last remaining
pebble has no predecessor and needs to be
irreversibly erased except in the initial block.
The initial pebble is put in front of the
lastly placed $n$th pebble which, having
done its duty as springboard for this
block, is subsequently
irreversibly erased. Therefore, the advancement of
each block requires two erasures, except the first
block which requires one, yielding a total
of $2m-1$ erasures. Let $G = \{ 1,2, \ldots, T_G \}$
be segmented as $B_1 b_1 \ldots B_m b_m$,
where each $B_i$ is a copy of interval $I_{n-1}$
above and $b_i$ is the node following $B_i$, for
$i=1, \ldots , m$. Hence, $T_G=m 2^{n-1}$.
We give the precise procedure in 
self-explanatory pseudo PASCAL using the procedures
given in the proof of Lemma~\ref{lem.bennett}.\\

\noindent
\begin{tabbing}
{\bf Procedure}  $A(n,m,G)$: \\ 
 {\bf for} \= $i:=1,2, \ldots ,m$: \\
 \> $F(n-1, B_i )$; \\
 \> erase pebble on node  $b_{i-1}$ ; \\
 \> put pebble on node  $b_i$ ; \\
 \>$F^{-1}(n-1, B_i )$ (removal of pebble from first
node of $B_i$ is an erasure) \\
\end{tabbing}
The simulation time $T'_G$ is 
$T'_G \approx 2m\cdot 3^{n-1} +2 
\approx 2m ( T_G/m)^{\log 3} = 2m^{1- \log 3 } T_G^{\log 3}$
for $T_G = m2^{n-1}$.
\end{proof}

\begin{theorem}[Space-Irreversibility Trade-off]
\label{theo.si}
(i) Pebble games $G$ of size $2^n-1$ can be won using $n$ pebbles
but not using $n-1$ pebbles.

(ii) 
If $G$ is a pebble game with a winning strategy
using $n$ pebbles without
erasures, then there is also a winning strategy for $G$
using $E$ erasures and $n-\log (E+1)$ pebbles (for $E$ is an odd
integer at least 1).

\end{theorem}
\begin{proof}
(i) By Lemmas~\ref{lem.bennett}, \ref{lem.pebble}.

(ii) By (i), $T_G = 2^n-1$
is the maximum length of a pebble game $G$
for which there is a winning strategy using $n$
pebbles and no erasures.
By Lemma~\ref{lem.erasure}, we can pebble a game $G$
of length $T_G= m2^{n-\log m}=2^n$ using $n+1-\log m$
pebbles and $2m-1$ erasures.
\end{proof}

We analyse the consequences of Theorem~\ref{theo.si}.
 It is convenient
to consider the special sequence of values
$E :=2^{k+2}-1$ for $k:=0,1, \ldots$.
Let $G$ be Bennett's pebble game of Lemma~\ref{lem.bennett}
of length $T_G=2^{n}-1$. 
It can be won using $n$ pebbles
without erasures, or using
$n-k $ pebbles plus $2^{k+2}-1$ erasures (which gives a gain
over not erasing as in Lemma~\ref{lem.bennett} only for $k \geq 1$), but not 
using $n-1$ pebbles. 

Therefore, we can exchange space use
for irreversible erasures.
Such a trade-off can be used to reduce
the excessive space requirements of the reversible simulation.
The correspondence between the
erasure pebble game and the
otherwise reversible computations
using irreversible erasures 
that if the pebble game uses $n-k$ pebbles
and $2^{k+2} -1$ erasures, then the otherwise reversible
computation uses $(n-k)S$ space and erases $(2^{k+2}-1)S$ bits
irreversibly. 

Therefore, a reversible simulation of an irreversible
computation of length $T=(2^n-1)S$ can be done using
$nS$ space using $(T/S)^{\log 3 } S$ time,
 but is impossible using $(n-1)S$ space. It can also
be performed using $(n-k)S$ space, $(2^{k+2}-1)S$
irreversible bit erasures and 
 $2^{(k+1)(1-\log 3 )+1} (T/S)^{\log 3} S$
time. In the extreme case
we use no space to store the history and erase about $4T$
bits. This corresponds to the fact that an irreversible
computation may overwrite its scanned symbol irreversibly
at each step.

\begin{definition}
\rm
Consider a simulation
using $S'$ storage space
and $T'$ time
which computes 
$y = \langle x, f(x) \rangle$ from $x$ in order to 
simulate
an irreversible computation
using $S$ storage space and $T$ time
which computes $f(x)$ from $x$.
The {\em irreversible simulation
cost} $B^{S'}(x, y)$ of the simulation
is the number of
irreversibly erased bits in the simulation (with the parameters $S,T,T'$
understood).
\end{definition}

If the irreversible 
simulated computation from $x$ to $f(x)$ uses $T$ steps, then for $S' = nS$ 
and $n =  \log (T/S)$ we have above treated the most space
parsimonious simulation which yields $B^{S'} (x,y) = 0$, with
$y= \langle x,f(x) \rangle$.

\begin{corollary}[Space-Irreversibility Trade-off]
Simulating a $T=(2^{n}-1)S$ step
irreversible computation from $x$ to $f(x)$ 
using $S$ space
 by
a computation from $x$ to $y = \langle x, f(x) \rangle$, the
irreversible simulation cost satisfies:

(i) $B^{(n- k)S } (x,y) \leq B^{nS}(x, y) + (2^{k+2}-1)S$,
for $n \geq k \geq 1$. 

(ii) $B^{(n-1)S}(x,y) > B^{nS}(x,y)$, for $n \geq 1$.

\end{corollary}

For the most space parsimonious
simulation with $n=\log (T/S)$ this means that
$B^{S(\log (T/S) - k) } (x,y) \leq 
B^{S \log (T/S)}(x, y) + (2^{k+2}-1)S$.

We {\em conjecture} that {\em all} reversible simulations
of an irreversible computation can
essentially be represented as the pebble game defined above,
and that consequently the lower bound of Lemma~\ref{lem.pebble}
applies to all reversible simulations of irreversible
computations. If this conjecture is true then
the trade-offs above turn into a space-irreversibility
hierarchy for polynomial time computations.

\section{Reversible Computation}
\label{sect.simulation}
Given that a computation is reversible, either by being
reversible {\em a priori} or by being a reversible simulation
of an irreversible computation, it will increasingly fill up the
memory with unwanted garbage information. Eventually this
garbage has to be irreversibly erased to create free memory space.
As before, the number of irreversibly
erased bits in
an otherwise reversible computation which replaces input $x$ by
output $y$, 
each unit counted as $kT \ln 2$,
represents energy dissipation.
Complementary to this idea, if such
a computation uses initially irreversibly provided bits
apart from input $x$, then they must be accounted at
the same negated cost as that for irreversible erasure.
Because of the reversibility
of the computation, we can argue by symmetry.
Namely, suppose we run a reversible computation starting when
memory contains input $x$ and additional record $p$,
and ending with memory containing output $y$ and additional
garbage bits $q$. Then $p$ is irreversibly provided, and $q$
is irreversibly deleted. But if we run the computation backward, then
the roles of $x,p$ and $y,q$ are simply interchanged.

Should we charge for the input $x$ or the output $y$? 
We do not
actually know where the input comes from, nor where the
the output goes to. Suppose we cut a computation
into two consecutive segments. If the output 
of one computation segment
is the input of another computation segment, then the
thermodynamic cost of the composition does not contain
costs related to these intermediate data. 
Thus, we want to measure just the number of irreversible 
bit operations
of a computation. We can view any computation
as consisting of a sequence of reversible and irreversible
operation executions. We want the irreversibility
cost to reflect all nonreversible parts of the computation.
The irreversibility cost of an otherwise reversible computation
must be therefore set to the {\em sum} of the number of
irreversibly provided and the number of irreversibly
erased bits.

We consider
the following axioms
as 
a formal basis on
which to develop a theory of
irreversibility of computation.
\begin{description}
\item[Axiom 1]
Reversible computations do not incur any cost.
\item[Axiom 2]
Irreversibly provided and irreversibly deleted bits
in a computation incur unit cost each.
\item[Axiom 3]
In a reversible computation which replaces input $x$ by
output $y$, the input $x$ is not irreversibly
provided and the output $y$ is not irreversibly
deleted.
\item[Axiom 4]
All physical computations are effective.
\end{description}

Axiom 4 is simply an extended form of {\em Church's Thesis}:
the notion of physical computation
coincides with effective 
computation which coincides with the formal notion
of Turing machines computation. Deutsch, \cite{De85},
and others have argued the possibility that this is false.
If that turns out to be the case then either our arguments
are to be restricted to those physical processes for which 
Axiom 4 holds, or, perhaps, one can extend the notion
of effective computations appropriately.

In reference \cite{BGLVZ93} we and others developed a 
theory of information distance with application
to the number of irreversible bit
operations in an otherwise reversible
computation. A precursor to this line of thought is \cite{Zu89a}.
  Among others,
they considered the information distance 
obtained by {\em minimizing} the
total amount of information flowing in and out during a reversible
computation in which the program is not retained. 

Since the ultimate limit
of energy dissipation by computation
is expressed in the number of bits in
the irreversibly erased
records,
we consider compactification of records.
Rather as in analogy of garbage collection by 
a garbage truck: the cost is less if we compact the garbage
before we throw it away.

The ultimate compactification of data which can be effectively
exploited is given by its Kolmogorov complexity.
This is a recursively invariant concept,
and expresses the limits to which effective methods can go.
Consequently, the mundane matter of energy dissipation
of physical computation can be
linked to, and expressed in, the pristine rigorous
notion of Kolmogorov complexity.

\subsection{Kolmogorov Complexity and Irreversibility Cost}
The Kolmogorov complexity, see \cite{LiVi93},
of $x$ is the length of the
{\em shortest} effective description of $x$.
Formally, this can be defined as follows.
Let $x,y,z \in {\cal N}$, where
${\cal N}$ denotes the natural
numbers and we identify
${\cal N}$ and $\{0,1\}^*$ according to the
correspondence $(0, \epsilon ), (1,0), (2,1), (3,00)$, $(4,01), \ldots$.
Hence, the length $|x|$ of $x$ is the number of bits
in the binary string $x$. 
Let $T_1 ,T_2 , \ldots$ be a standard enumeration
of all Turing machines. Without
loss of generality we assume that all machines in this 
paper have binary input, storage, and output.
Consider a standard reversible mapping that maps
a pair of integers $x,y$ to another integer
$\langle x ,y \rangle$. 
Similarly,
$\langle  \langle x , y \rangle , z \rangle$
reversibly maps {\em triplets} of integers to
a single integer. Let the mapping be Turing-computable.

\begin{definition}
Let $U$ be an appropriate universal Turing machine
such that $U(\langle \langle i,p \rangle ,y \rangle ) =
T_i (\langle p,y\rangle)$ for all $i$ and $\langle p,y\rangle$.
The {\em Kolmogorov complexity} of $x$ given $y$ (for 
free) is
\[C(x|y) = \min\{|p|: U (\langle p,y\rangle)=x , p \in \{0,1\}^*, i
\in {\cal N} \}. \]
\end{definition}
Axioms 1---4 lead to the definition of the irreversibility cost
of a computation as
the number of bits we added plus the number of
bits we erased in computing one string from another.
Let ${\bf R} = R_1 , R_2 , \ldots$ be a standard
enumeration of reversible Turing machines, \cite{Be73}.

The irreversibility cost of otherwise reversibly computing
from $x$ to $y$ is the
number of extra bits (apart from $x$) that must be irreversibly
supplied at the beginning, plus the number of garbage bits (apart from
$y$) that must be irreversibly erased at the end of the computation to
obtain a `clean' $y$.
  The use of irreversibility resources
in a computation is expressed in terms of
this cost, which is one of the information distances
considered in \cite{BGLVZ93}. It is shown to be
within a logarithmic additive term of the sum of
the conditional complexities, $C(y|x)+C(x|y)$.

\begin{definition}\label{def.icost}
The {\em irreversibility cost} 
$E_R (x,y)$ of computing $y$ from $x$
by a reversible Turing machine $R$ is
is
\[ E_R (x,y) = \min \{|p|+ |q|: R (\langle x,p\rangle)=\langle y,q\rangle \}. \]
We denote the class of all such cost functions
by ${\cal E}$.
\end{definition}

We call an element $E_Q$ of ${\cal E}$ a 
{\em universal irreversibility cost function},
if $Q \in {\bf R}$, and for all $R$ in {\bf R}
\[ E_{Q} (x,y) \leq E_{R} (x,y) + c_{R} ,\]
for all $x$ and $y$, where $c_R$ is a constant which
depends on $R$ but not on $x$ or $y$.
Standard arguments from the theory of Turing machines
show the following.

\begin{lemma}
There is a universal irreversibility cost function
in ${\cal E}$. Denote it by $E_{UR}$.
\end{lemma}
\begin{proof}
In \cite{Be73} a universal reversible Turing machine $UR$
is constructed which satisfies the optimality requirement.
\end{proof}

Two such universal (or optimal) machines $UR$ and $UR'$ will
assign the same irreversibility cost to a computation
apart from an additive constant term $c$ which is {\em independent}
of $x$ and $y$
(but does depend on $UR$ and $UR'$).
We select a reference universal function $UR$
and define the
{\em irreversibility cost} $E(x,y)$ of
computing $y$ from $x$ as
\[E(x,y) \equiv  E_{UR} (x,y) . \]

In physical terms
this cost is in units of $kT \ln 2$, where $k$ is Boltzmann's constant,
$T$ is the absolute temperature in degrees Kelvin,
and $\ln$ is the natural logarithm.

Because the computation is reversible, this definition
is {\em symmetric}: we have $E(x,y)=E(y,x)$.

In our definitions we have pushed all bits to be 
irreversibly provided to the start of the computation
and all bits to be erased to the end of the computation.
It is easy to see that this is no restriction. If we have
a computation where irreversible acts happen throughout
the computation, then we can always mark the bits to
be erased, waiting with actual erasure until the end
of the computation.
Similarly, the bits to be provided can be provided
(marked) at the start of the computation while the actual
reading of them (simultaneously unmarking them)
takes place throughout the computation).

\subsection{Computing Between $x$ and $y$}

Consider a general computation which
outputs string $y$ from input string $x$.
We want to know the
minimum irreversibility cost for such computation. 
The result below appears in \cite{BGLVZ93} with a
different proof.

\begin{theorem}[Fundamental theorem]\label{thermo.distance}
Up to an additive logarithmic term\footnote{Which is
$O(\min \{C(C(y|x)|y),C(C(x|y)|x) \})=
O( \log \min \{C(y|x),C(x|y)\})$.
It has been shown, \cite{Ga74},
that for some $x$ of each length $n$
we have \[ \log n - \log \log n \leq C(C(x)|x),\]
and for all $x$ of length $n$ we have
\[ C(C(x)|x) \leq \log n + 2 \log \log n .\]},
\[E(x,y) = C(x|y)+C(y|x) .\]
\end{theorem}

\begin{proof}
We prove first an upper bound and then a lower bound.
\begin{claim}\label{upperbound}
$E(x,y) \leq C(y|x)+C(x|y)+2[C(C(y|x)|y)+C(C(x|y)|x)]$.
\end{claim}
\begin{proof}
We start out the computation with 
programs $p,q,r$. Program $p$ computes
$y$ from $x$ and $|p| =C(y|x)$. Program $q$ computes
the value $C(x|y)$ from $x$ and $|q|=C(C(x|y)|x)$.
Program $r$ computes the value $C(y|x)$ from $y$ and
$|r| =C(C(y|x)|y)$. To separate the different binary
programs we have to encode delimiters. This takes
an extra additional number of bits logarithmic
in the two smallest length of elements $p,q,r$.
This extra log term is absorbed in the additive log
term in the statement of the theorem.
The computation is as follows. Everything is executed reversibly
apart from the final irreversible erasure.
\begin{enumerate}
\item
Use $p$ to compute $y$ from $x$ producing garbage bits $g(x,y)$.
\item
Copy $y$, and use one copy of $y$ and $g(x,y)$ to reverse
the computation to $x$ and $p$. Now we have $p,q,r,x,y$.
\item\label{s.1}
Copy $x$, and use one copy of $x$ and $q$ to compute
$C(x|y)$ plus garbage bits.
\item\label{s.2}
Use $x,y,C(x|y)$ to dovetail the running of all programs
of length $C(x|y)$ to find $s$, a shortest program
to compute $x$ from $y$. Doing this, we produce more garbage bits.
\item
Copy $s$, and reverse the computations in Steps~\ref{s.2}, \ref{s.1},
canceling the extra copies and all garbage bits.
Now we have $p,q,r,s,x,y$.
\item\label{p.1}
Copy $y$, and use this copy to compute the value $C(y|x)$
from $r$ and $y$ producing garbage bits.
\item\label{p.2}
Use $x,y,C(y|x)$,
to dovetail the running of all programs of length
$C(y|x)$ to obtain a copy of $p$, the shortest program
to compute $y$ from $x$, producing more garbage bits. 
\item
Delete a copy of $p$ and reverse the computation of
Steps~\ref{p.2}, \ref{p.1} canceling the superfluous copy of $y$
and all garbage bits.
Now we are left with $x,y,r,s,q$.
\item
Compute from $y$ and $s$ a copy of $x$ and cancel a copy of $x$.
Reverse the computation. Now we have $y,r,s,q$.
\item
Erase $s,r,q$ irreversibly.
\end{enumerate}
We started out with additional shortest programs $p,q,r$
apart from $x$. We have
irreversibly erased the shortest
programs $s,q,r$, where $|s| =C(x|y)$, leaving only $y$.
This proves the claim.
\end{proof}

Note that all bits supplied in the beginning to the computation,
apart from input $x$,
as well as all bits irreversibly erased
at the end of the computation, 
are {\em random} bits. This is because we supply and delete
only shortest programs, and a shortest program $p$
satisfies $C(p) \geq |p|$, that is, it is maximally random.

\begin{claim}\label{lowerbound}
$E(x,y) \geq C(y|x)+C(x|y)$.
\end{claim}
\begin{proof}
To compute $y$ from $x$ we must be given a program to do so
to start out with. By definition the shortest
such program has length $C(y|x)$.

Assume the computation from $x$ to $y$ produces $g(x,y)$
garbage bits.
Since the computation
is reversible we can compute 
$x$ from $y$ and $g(x,y)$.
Consequently, $|g(x,y)| \geq C(x|y)$ by definition~\cite{Zu89a}.
To end the computation with $y$ alone we therefore
must irreversibly erase $g(x,y)$ which is
at least $C(x|y)$ bits.
\end{proof}

Together Claims~\ref{upperbound}, \ref{lowerbound}
prove the theorem.
\end{proof}

Erasing a record $x$ is actually a computation from
$x$ to the empty string $\epsilon$. Hence its irreversibility
cost
is $E(x, \epsilon )$, and given by a corollary to
Theorem~\ref{thermo.distance}.

\begin{corollary}\label{energy.coro}
Up to a logarithmic additive term, 
the irreversible
cost of erasure is
$E(x, \epsilon )= C(x)$.
\end{corollary}

\section{Trading Time and Space for Energy}

In order to erase a record $x$, 
Corollary~\ref{energy.coro} actually requires us to have,
apart from $x$, a program $p$ of length $C(C(x)|x)$
for computing $C(x)$, given $x$. The precise bounds
are $C(x) \leq E(x, \epsilon ) \leq C(x)+2C(C(x)|x)$.
This optimum is not effective, it requires that $p$
be given in some way. But we can use the same
method as in the proof of Theorem~\ref{thermo.distance},
by compressing $x$ using some time bound $t$. Using
space bounds is entirely analogous.
Instead of the superscript `$t$',
we can use everywhere `$s$', where `$s(\cdot )$'
denotes a space bound, or `$t,s$' to denote simultaneous
time and space bounds.

First we need some definitions as in \cite{LiVi93}, page 378
and further. 
Because now the time bounds are
important we consider the universal Turing machine $U$
to be the machine with two work tapes which
can simulate $t$ steps of a multitape Turing machine $T$
in $O(t \log t)$ steps.
If some multitape Turing machine $T$
computes $x$ in time $t$ from a program $p$,
then $U$ computes $x$ in time $O(t \log t)$ from
$p$ plus a description of $T$. 
\begin{definition}
Let $C^t(x|y)$ be the {\em minimal length} of binary program
(not necessarily reversibly) for the 
two work tape universal Turing machine $U$
computing $x$ given $y$ (for free) {\em in time} $t$. Formally,
\[ C^t (x|y) = \min_{p \in {\cal N}}
 \{|p|: U(\langle  p,y \rangle )= x
\mbox{ in $\leq t(|x|)$ steps} \}. \]
$C^t(x|y)$ is called the $t$-{\em time-limited conditional Kolmogorov
complexity} of $x$ given $y$. The unconditional
version is defined as $C^t(x):=C^t(x, \epsilon)$.
A program $p$ such that $U (p)=x$
in $\leq t(|x|)$ steps and $|p|=C^t(x)$ is denoted as $x^*_t$.
\end{definition}

Note that with $C_T^t (x|y)$ the
conditional $t$-time-limited Kolmogorov complexity
with respect to Turing machine $T$, for all $x,y$,
$C^{t'} (x|y) \leq C_T^t (x|y) + c_T$, where $t'=O(t \log t)$ and
$c_T$ is a constant depending on $T$ but not on $x$
and $y$.

This $C^t(\cdot)$ is the standard definition of time-limited
Kolmogorov complexity.
However, in the remainder of the paper
we always need to use reversible computations. Fortunately,
in \cite{Be89} the following is shown (using the simulations
refered to in Section~\ref{sect.simulation}).
\begin{lemma}\label{lemma.Bennett}
For any $\epsilon >0$, ordinary multitape Turing machines
using $T$ time and $S$ space can be simulated by reversible
ones using time $O(T)$ and space $O(ST^{\epsilon})$
(or in $O(T)$ time and space $O(S+T)$).
\end{lemma}
To do effective erasure of compacted information,
we must at the start of the computation provide
a time bound $t$. Typically, $t$ is a recursive function
and the complexity of its description is small, say $O(1)$.
However, in Theorem~\ref{time.1} we allow for very large
running times in order to obtain smaller $C^t(\cdot)$
values.
(In the theorem below $t$ need not necessarily be
a recursive function
$t(|x|)$, but can also be used nonuniformly. This 
leads to a stronger result.)
\begin{theorem}[Irreversibility cost of effective erasure]\label{time.1}
If $t(|x|) \geq |x|$ is a
time bound which is provided at the start of the
computation,
then erasing an $n$ bit record $x$ by an otherwise
reversible computation can be done
in time (number of steps) $O(2^{|x|}t(|x|))$
at irreversibility cost
$C^t (x)+ 2C^t (t|x)+4 \log C^t (t|x)$ bits. (Typically we 
consider $t$ as some standard explicit time bound
and the last two terms adding up to $O(1)$.)
\end{theorem}
\begin{proof}
Initially we have in memory
input $x$ and a program $p$ of length $C^t(t,x)$
to compute
reversibly $t$ from $x$. To separate
binary $x$ and binary $p$ we need to encode a delimiter
in at most $2 \log C^t (t|x)$ bits.
\begin{enumerate}
\item
Use $x$ and $p$ to reversibly compute $t$. 
Copy $t$ and reverse the computation.
Now we have $x$, $p$ and $t$.
\item
Use $t$ to reversibly dovetail the running
of all programs of length less
than $x$ to find the shortest one halting
in time $t$ with output $x$.
This is $x^*_t$. The computation has produced
garbage bits $g(x,x^*_t)$.
Copy $x^*_t$, and reverse the computation to obtain $x$
erasing all garbage bits $g(x,x^*_t)$.
Now we have $x,p,x^*_t,t$ in memory.
\item 
Reversibly compute $t$ from $x$ by $p$, cancel one copy of $t$,
and reverse the computation. Now we have $x,p,x^*_t$
in memory.
\item
Reversibly cancel $x$ using $x^*_t$ by the standard method,
and then erase $x^*_t$ and $p$ irreversibly. 
\end{enumerate}
\end{proof}


\begin{corollary}\label{cor.Kt}
The irreversibility cost satisfies
\[ E(x, \epsilon) \geq \lim_{t 
\rightarrow \infty} C^{t} (x) = C(x) , \]
and by Theorem~\ref{thermo.distance} 
up to an additional logarithmic term
\[ E(x, \epsilon)
= C(x). \]
\end{corollary}

Essentially, by spending more time we can
reduce the thermodynamic cost of erasure of $x^*_t$ to its
absolute minimum. In the limit we spend
the optimal value $C(x)$ by erasing $x^*$,
since $\lim_{t \rightarrow \infty} x^*_{t} = x^*$.
This suggests the existence of
a trade-off hierarchy between time and
energy. The longer 
one reversibly computes on a particular given string 
to perform final irreversible
erasures, the less bits are erased and energy is dissipated.
This intuitive assertion
will be formally stated and rigourously proved below as
Theorem~\ref{theorem.hierarchy}:
for each length $n$
we will construct a particular string which can be compressed
more and more by a sequence of about $\sqrt{n}/2$ growing
time bounds.
We proceed through a sequence
of related `irreversibility' results.

\begin{definition}\label{def.tle}
Let $UR$ be the reversible version of the two worktape
universal Turing machine, simulating the latter in
linear time by Lemma~\ref{lemma.Bennett}.
Let $E^t(x,y)$ be the {\em minimum irreversibility cost} of
an otherwise reversible computation from 
$x$ to $y$ {\em in time} $t$. Formally,
\[ E^t (x,y) = \min_{p,q \in {\cal N}}
 \{|p|+|q|: UR(\langle x,p\rangle)=\langle y,q\rangle 
\mbox{ in $\leq t(|x|)$ steps} \}. \]
\end{definition}

Because of the similarity with Corollary~\ref{cor.Kt} ($E(x, \epsilon)$
is about $C(x)$) one is erroneously led to believe that
$E^t(x, \epsilon) = C^t(x)$ up to a log additive term.
However, the time-bounds introduce many differences.
To reversibly compute $x^*_t$ we may require (because of the
halting problem) at least $O(2^{|x|} t(|x|))$ steps after having
decoded $t$, as indeed is the case in the proof of Theorem~\ref{time.1}.
In contrast, $E^t(x, \epsilon)$ is about
the number of bits erased in an otherwise
reversible computation which uses
at most $t$ steps. Therefore, as far as we know possibly
$C^t (x) \geq E^{t'} (x, \epsilon)$ 
implies $t' = \Omega ( 2^{|x|}t(|x|))$.
More concretely, it is easy to see that for each 
$x$ and $t(|x|) \geq |x|$,
\begin{equation}\label{eq.CE}
E^t(x,\epsilon ) \geq C^t(x) \geq  E^{t'} (x, \epsilon)/2 , 
\end{equation}
with $t'(|x|) = O(t(|x|)$.
Namely, the left inequality follows since $E^t (x, \epsilon)$
means that we can reversibly compute from $\langle x,p \rangle$
to $\langle \epsilon , q \rangle$ in $t(|x|)$ time where
$|p|+|q|= E^t(x, \epsilon)$. But this means that we can compute
$x$ from $q$ in $t(|x|)$ time (reversing the computation)
and therefore $C^t(x) \leq |q|$.
The right inequality follows by the following scenario.
At the start of the computation provide apart from input
$x$ also (irreversibly) $x^*_t$, 
the shortest binary program computing $x$
in at most $t(|x|)$ steps, so $|x^*_t|=C^t(x)$. 
From $x^*_t$ reversibly compute a copy of 
$x$ in $O(t(|x|))$ time, Lemma~\ref{lemma.Bennett},
cancel the input copy of $x$,
reverse the computation to obtain $x^*_t$ again,
and irreversibly erase $x^*_t$.  

Theorem~\ref{time.1} can be restated in terms of $E^t(\cdot)$
as
\[E^{t'} (x, \epsilon) \leq C^t (x)+ 2C^t (t|x)+4 \log C^t (t|x),\]
with $t'(|x|)=O(2^{|x|}t(|x|))$.
Comparing this to the righthand inequality of Equation~\ref{eq.CE}
we have improved the upper bound on erasure cost at the expense of
increasing erasure time. However, these bounds only suggest
but do not actually prove that we 
can exchange irreversibility for time.
Below, we establish rigorous
time-space-irreversibility
trade-offs.

\section{Trade-off  Hierarchy}
The following result establishes the existence
of a trade-off hierarchy of time versus irreversibility
for exponential time computations. 
\footnote{A superficially similar but quite different result for 
the time-limited so-called {\em uniform} Kolmogorov
complexity variant $C(x;|x|)$ was 
given in \cite{Da73b}, but is too weak for our purpose. 
There the time bound $t$
denotes {\em decompression} time while in $E^{t'}(x, \epsilon)$
the time bound $t'$ relates 
to {\em compression} time. Moreover, the result shows
a hierarchy in the sense 
that for certain classes of unbounded functions 
$\{ f_i : i \in {\cal N}\}$ (satisfying $2 f_{i+1}(n) \leq f_{i} (n)$),
there exists a recursive infinite sequence $\omega_1  \omega_2 \ldots$
and a recursive sequence of time bounds 
$\{ t_i : i \in {\cal N}\}$, such that for each $i \geq 1$ there
are infinitely many $n$ such that 
$C^{t_i} (\omega_1  \dots  \omega_n ; n) > f_i (n)$ while for all $n$
we have $C^{t_{i+1}} (\omega_1  \dots  \omega_n ; n) \leq f_{i} (n)$.
See also Exercise 7.7 in \cite{LiVi93}.
Note that the set of infinitely many $n$ in the statement
above may constitute
a different disjoint set
for each $i$. Hence, for each pair of distinct time bounds there
are initial segments of
the {\em single infinite} sequence which exhibit different 
compressions, but not necessarily the same initial segment
exhibiting pairwise different compressions for more than two
time bounds simultaneously, 
let alone 
a $\sqrt{n}/2$ level time-erasure hierarchy
for {\em single finite} sequences of {\em each} length $n$ as in 
Theorem~\ref{theorem.hierarchy}. Even if it could be shown
that there are infinitely many initial segments,
each of which exhibits 
maximally many pairwise different compressions for different
time bounds, it would
still only result in a $\log n$ level time-decompression
hierarchy for sequences of infinitely many lengths $n$.
In contrast, the proof of Theorem~\ref{theorem.hierarchy}
also yields the analogous $\sqrt{n}/2$ level
time-decompression hierarchy
for Kolmogorov complexity.}
The proof proceeds by a sequence of diagonalizations which
just fit in the exponential time bounds.

\begin{theorem}[Irreversibility-time trade-off hierarchy]
\label{theorem.hierarchy}
For every large enough $n$
there is a string $x$ of length $n$ and a sequence of $m=\frac{1}{2}
\sqrt n$
time functions $t_1(n) < t_2(n) <  \ldots < t_m(n)$, 
such that
\[ 
E^{t_1}(x,\epsilon ) > E^{t_2}(x,\epsilon ) > \ldots > E^{t_m}(x, \epsilon ).
\]
\end{theorem}

\begin{proof}
Given $n$, we will construct a string $x$ of length $n$
satisfying the requirements
of the theorem. String $x$ will be constructed in $m$ steps, and 
$x$ will contain $m$ blocks $x_1,x_2, \ldots ,x_m$ 
each of length $b= n/m$. The idea is
to make these blocks harder and harder to compress. Define, 
for $1 \leq k \leq m$, 
\[ t_k(n) = 2^{kn} .\]
In our construction, we will enforce the following things:
\begin{itemize}
\item[$\bullet$]
All $m$ blocks can be compressed iff given enough time.
Precisely, $x_k$ can be compressed to $O(\log n)$ size 
given $t_{k+1}(n)$
time, but given $t_k (n)$ time $x_k$ cannot be compressed at all.
\item[$\bullet$]
No ``collective compression''. If $x_k$ cannot be compressed
in time $t$ then the concatenation $x_k \ldots x_m$, 
as a single string,
cannot be compressed in time $t$ either. In the construction,
we will use only prefixes from strings in set $S_k$ which
consists of strings that are not compressible in time $t_k(n)$.
\end{itemize}

\subsection*{Algorithm to Construct $x$}
\begin{description}
\item[Initialize:]
\subitem
Set $S_0 :=\{0,1\}^n$, the set of all strings of length $n$, 
and $t_0(n):=0$ and $k:=0$.
\item [Repeat For $k+1:= 1, \ldots,m$:]
\subitem
$/*$ Starting the $(k+1)$st repetition,
the first $k$ blocks $x_1, \ldots , x_k$ 
of $x$ have already been constructed and in the $k$th
repetition we have 
constructed a set $S_k$ consisting of strings of length
$n-kb$, no element of  which can be computed from programs of
length less than $n-kb -2k$ in time $t_k (n)$. Furthermore,
\[ 2^{n-kb} \geq |S_k | \geq 2^{n- kb -2k}. \; \; */ \]

\subitem
Construct $x_{k+1}$ from $S_k$ as follows. 
Let $s$ be the lexicographic first string of length $b$ such that 
\begin{equation}\label{thermo1}
|\{ s' : ss' \in S_k \}| \geq 2^{n-(k+1)b-2k} .
\end{equation}
Such a $s$ exists by Claim~\ref{trade.claim.1}.
Set $x_{k+1} := s$.
\subitem
Construct $S_{k+1}$ from $S_k$ and $x_{k+1}$ as follows. 
Let $S_k' = \{ s' : x_{k+1}s' \in S_k \}$. We have 
$|S_k'| \geq 2^{n-(k+1)b-2k}$ by Equation~\ref{thermo1}.
Simulate each of the programs of length less than $n-(k+1)b - 2(k+1)$
for $t_{k+1}(n)/2$ steps. Set $S_{k+1}$
to be the set of all strings $s'$ of length $n-(k+1)b$ such that
$s' \in S_k'$ and $s'$ is not an output of any
of the above simulations.
We have $|S_{k+1} | \geq 2^{n-(k+1)b-2(k+1)}$. Trivially,
$2^{n-(k+1)b} \geq |S_{k+1}|$.
This finishes the description of the algorithm.
\end{description}


\begin{claim}\label{trade.claim.1}
There is a string $s$ of length $b$ such that 
\[
| \{ s' : ss' \in S_k \} | \geq 2^{n-(k+1)b-2k} .
\]
\end{claim}
\begin{proof}
If the claim is false, then
the number of elements in $S_k$ must be less than
\[ 2^b 2^{n-(k+1)b-2k} = 2^{n-kb-2k} , \]
which is a contradiction.
\end{proof}


\begin{claim}\label{thermo.claim4}
For each $k=1, \ldots , m$, the sequence of
blocks $x_1, \ldots ,x_k$ can be computed
by a $O(\log n)$ sized program in time $t_{k+1}(n)/n$.
\end{claim}
\begin{proof}
Using the values of
$n,b,k$ and a constant size
program  we can execute the Construction algorithm
up to and including the $(k-1)$th repetition in at most
\begin{eqnarray*}
\sum_{i=1}^{k-1} 2^{n-ib-2i} t_i(n) & \leq & 
2^{n-b-2} \sum_{i=1}^{k-1} 2^{ni} \\
 & \leq & 2^{n-2\sqrt n -2} 2^{n(k-1)+1} 
\leq 2^{nk}/2n = t_k(n)/2n 
\end{eqnarray*}
steps.
Subsequently, we can find $x_k$ in at most
$n |S_{k-1}| \leq t_k(n)/2n$ steps.
Therefore, in a total number of steps
not exceeding $t_{k}(n)/n$,
we can compute the list $x_1, \ldots, x_k$ by a $O(\log n)$ 
size program.
\end{proof}

\begin{claim}\label{energy.eq1}
Let $n,b,m,k$ be as above. Then,
$E^{t_{k}} (x, \epsilon ) \leq n-kb + O(\log n)$.
\end{claim}

\begin{proof}
Using Claim~\ref{thermo.claim4},
we can compute $x$ from an $O(\log n)$
bits program and $x_{k+1} , \ldots x_m$
($\leq n - kb + O(\log n)$ bits),
collectively denoted as program $p$, in $t_k(n)/n$ time.
Trivially, we can compress $x$ using an a program
$q$ (containing $n,m,k$) with $|q| = O(\log n)$ to $p$ in
$t_k(n)/n$ time.
Using methods developed earlier in this paper,
we can erase $x$ in an
otherwise reversible computation
irreversibly erasing only $|p|=n-kb + O(\log n)$ bits and
irreversibly providing only $|q|$ bits, in
$t_{k} (n)$ time, as follows. By Lemma~\ref{lemma.Bennett}
the overhead incurred by making these computations
reversible is only linear.
\begin{enumerate}
\item
Reversibly compute $p$ from $x$ and $q$, with 
garbage $g(x,p)$, using $O(t_k(n)/n)$ steps.
Now we have $p, g(x,p)$.
\item
Copy $p$, then reverse the computation of Item 1, absorbing
the garbage bits $g(x,p)$, using at most $O(t_k(n)/n)$ steps.
Now we have $x,p,q$.
\item
Reversibly compute from $p$ to $x$, with garbage
$g(p, x)$; then cancel a copy of $x$, 
using at most $O(t_k(n)/n)$ time.
Now we have $x,q,g(p,x)$.
\item
Reverse the computation of Item 3, absorbing the garbage bits 
$g(p,x)$, leaving only $p,q$,
then remove $p$ and $q$ irreversibly, 
using at most time $t_k(n)/n)$.
\end{enumerate}
In total, above erasing procedure uses $O(t_k (n)/n)$ steps and
erases $|p| + |q|$ bits irreversibly and provides $|q|$
bits irreversibly. This proves the claim.
\end{proof}

\begin{claim}\label{energy.eq2}
Let $n,b,m,k$ be as above. Then,
$E^{t_k} (x, \epsilon ) \geq n-kb - 2k - 7 \log n$.
\end{claim}
\begin{proof}
Suppose the contrary, and we can reversibly
compute $\langle \epsilon , q \rangle$ from $\langle x,p \rangle$,
with 
\[ |q| \leq E^{t_{k}} (x, \epsilon) < n - kb -2k -7 \log n.\]
Then, reversing the computation, in $t_k (n)$ time
a program $q$ of size at most $n-kb-2k -7 \log n$ can
reversibly compute $x$ possibly together with (here
irrelevant) garbage $p$.
Therefore, this program $q$ plus descriptions of $n,m,k$
of total size at most 
$n-kb - 2k -  \log n$ can (possibly non-reversible) compute
$x_{k+1} \ldots x_m$ in $S_{k}$ in time $t_{k}(n)$. 
But this contradicts
the definition that no string in $S_{k}$ can be
(non-reversible) computed in time
$t_{k} (n)$ by a program of less than 
$n-kb - 2k$ bits.
\end{proof}

By Claim~\ref{energy.eq1} using $(k+1)$ for $k$,
Claim \ref{energy.eq2}, and
the assumption that $b = 2 \sqrt n $, we have 
for all $k$ such that $1 \leq k < m$,
\[ E^{t_k} (x,\epsilon ) > E^{t_{k+1}} (x,\epsilon ). \]
The theorem is proven.
\end{proof}

We have demonstrated our theorem for the case when $y=\epsilon$.
For $y \neq \epsilon$, it is easy to see that the proof still
holds if we simply require that $|x_k| \geq |y|^2$ for each $k$
and make sure $y$ is always an extra input when we simulate
all the short programs to construct $x$.
Therefore, the theorem
can be generalized to the following.
\begin{corollary}\label{cor.thier}
For every $y$ and 
every large enough $n$ 
there is a string $x$ of length $n$ and a sequence of 
$m= \frac{1}{2} \sqrt n$
time functions $t_1(n) < t_2(n) <  \ldots < t_m(n)$, 
such that
\[ 
E^{t_1}(x,y ) > E^{t_2}(x,y ) > \ldots > E^{t_m}(x, y ).
\]
\end{corollary}

Various different information distances
and thermodynamic cost measures can be considered.
For example, considering only the maximum
of the irreversibly provided bits or initial program
and the irreversibly erased bits or final garbage.
Following Landauer, \cite{La61}, we may for the 
energy-dissipation consider only the number of irreversibly
erased bits. All such measures and also time-limited
Kolmogorov complexities exhibit the same or very similar
time-irreversibility trade-offs by the above proof.
The result is common to all reasonable cost measures,
and the reader is referred to \cite{BGLVZ93}
for the fine distictions among them and for their physical
meanings.

\section{Extreme Trade-offs}
While the time functions in Theorem~\ref{theorem.hierarchy}
are much too large for practical computations,
they are much smaller than the times required to squeeze
the irreversibility out of those computations
most resistant to being made reversible.
The following %
{\it blow-up}
Lemma \ref{theo.barzdin}, \cite{Ba68},
was one of the very first results
in `time-limited' Kolmogorov complexity. 
\begin{definition}
\rm
Let set $A \subseteq {\cal N}$. 
Its {\em characteristic sequence} $\chi = \chi_1 \chi_2 \ldots$ 
is defined by $\chi_i =1$
if $i \in A$ and 0 otherwise (all $i \in {\cal N}$).
If $A$ is recursively enumerable
(r.e. for short), then we call $\chi$ an {\em r.e. sequence}.
\end{definition}

\begin{lemma}\label{theo.barzdin}

(i) There is an r.e. sequence $\chi$ such that
for each total recursive function $t$ there is a constant $c_t$
($0< c_t <1$), such that
for each $n$ we have $C^t ( \chi_1 \ldots \chi_n |n) \geq  c_t n$.




(ii) 
  Each r.e. sequence $\chi$ satisfies
  $C( \chi_1 \ldots \chi_n )  \leq  2 \log n +c $ 
for all $n$, where $c $ is a
 constant dependent on $\chi$ $($but not on $n )$.
  

\end{lemma}

It follows from Equation~\ref{eq.CE} that 
$E^t(x,\epsilon) \geq C^t (x) $ for all time bounds $t$.
Then, by  Lemma~\ref{theo.barzdin} (i),
there is a sequence $\chi=\chi_1 \chi_2 \ldots $
such that for each total recursive 
time bound $t$ there is a constant $c_t > 0$
such that $ E^t (\chi_1 \ldots \chi_n, \epsilon ) > c_t n $.

However, for a large enough nonrecursive
time bound $T$ (like $T(n) = \infty$) we have
$E^{T}( \chi_1 \ldots \chi_n) = C ( \chi_1 \ldots \chi_n )$,
for all $n$.
Then, by Lemma~\ref{theo.barzdin} (ii)
all such sequences $\chi=\chi_1 \chi_2 \ldots $
satisfy  $E^{T}( \chi_1 \ldots \chi_n) \leq 2 \log n + c$,
for all $n$ (with $c>0$ a constant depending only on $\chi$).
These two facts together
demonstrate that with respect to the irreversible erasure
of certain strings 
exponential energy dissipation savings
are sometimes possible when any recursive time bound
whatsoever available for the erasure procedure
is changed
to a 
large enough nonrecursive time bound.

\begin{theorem}
There is a r.e. sequence $\chi$ 
and some (nonrecursively) large time bound $T$,
such that for each total recursive time bound $t$,
for each initial segment $x$ of $\chi$
\[ E^t (x, \epsilon ) > c_t 2^{E^{T}(x, \epsilon )/2} ,\]
where $c_t > 0$ is a constant depending only on $t$ and $\chi$.
\end{theorem}

The trade-off can be slightly improved 
for a restricted set of infinitely
many initial segments of $\chi$ in the sense of dropping the
dependency of the constant $c_t$ on $t$.
Using a result \cite{Da73}, page 306 last line,
instead of Barzdin's Lemma~\ref{theo.barzdin} (i),
 changes the theorem to:

 ``There is an r.e. sequence $\chi$
and some  (nonrecursively) large time bound $T$,
  such that
for each total recursive time bound $t$,
for infinitely many initial segments $x$ of $\chi$:
  \[ E^t (x, \epsilon ) > c 2^{E^{T}(x, \epsilon )/2} ,\]
where $c$ is a constant depending only on $\chi$.''

In  other situations  the trade-off can be even 
more extreme. We just mention the results and
do not explain the esotheric notions involved but refer
the interested reader to the cited literature.
For so-called Mises-Wald-Church random binary sequences 
$\omega = \omega_1 \omega_2 \ldots$
where the admissible place-selection rules are restricted
to the total recursive functions (instead of the more common
definition using the partial recursive functions) Daley 
has
shown the following. (We express his
results in the Kolmogorov complexity variant 
called {\em uniform complexity}
he uses. In \cite{LiVi93}, Exercise 2.42, the
uniform complexity of $x$ is denoted as $C(x;l(x))$) 

There are sequences $\omega$ as described above such that
for each unbounded total recursive function $f$ 
(no matter how small)
we have 
$C(\omega_1 \ldots \omega_n;n) < f(n)$ for all large enough $n$, \cite{Da75}, 
given as  Exercise 2.47 Item (c) in \cite{LiVi93}.

Moreover, for all such $\omega$
and each total unbounded nondecreasing time bound $t$
(no matter how great)
there are infinitely many $n$ such that 
$C^t (\omega_1 \ldots \omega_n;n ) \geq n/2$, \cite{Da73},
given as Exercise 7.6 in
  \cite{LiVi93}.

Defining a uniform energy dissipation variant $E_u (\cdot, \cdot)$
similar to Definitions~\ref{def.icost}, \ref{def.tle} but
using the uniform Kolmogorov complexity variant,
these results translate in the now familiar way to the
statement that the energy-dissipation can be reduced 
arbitrarily computably far by using enough (that is, a noncomputable
amount of) time.

\begin{lemma}
There is a sequence $\omega$ and
a (nonrecursively) large time bound $T$,
such that
 for each unbounded total recursive function $f$,
no matter how large, 
for each total recursive time bound $t$,
there are infinitely many $n$ for which
  \[ E_u^t ( \omega_1 \ldots \omega_n , \epsilon ) 
> f(E_u^{T}( \omega_1 \ldots \omega_n , \epsilon )) .\]
\end{lemma}

\section*{Acknowledgements}
We thank the referees for their valuable comments
of how to improve presentation, and we thank one
of the referees for the interesting idea to investigate
reversible simulations and extreme trade-offs.

\bibliographystyle{plain}

\begin{thebibliography}{99}

\bibitem[Barzdin', 1968]{Ba68}
Y.M. Barzdin', Complexity of programs to determine whether 
natural numbers not
greater than $n$ belong to a recursively enumerable set,
{\em Soviet Math. Dokl.},9 (1968), 1251-1254.

\bibitem[Bennett, 1973]{Be73}
C.H. Bennett.
\newblock Logical reversibility of computation.
\newblock {\em IBM J. Res. Develop.}, 17:525--532, 1973.

\bibitem[Bennett, 1982]{Be82}
C.H. Bennett.
\newblock The thermodynamics of computation---a review.
\newblock {\em Int. J. Theoret. Phys.}, 21(1982), 905-940.

\bibitem[Bennett, 1989]{Be89}
C.H. Bennett.
\newblock Time-space trade-offs for reversible computation.
\newblock {\em SIAM J. Comput.}, 18(1989), 766-776.

\bibitem[Bennett {\em et al.,} 1993]{BGLVZ93}
C.H. Bennett, P. G\'acs, M. Li, P.M.B. Vit\'anyi and W.H Zurek,
\newblock Thermodynamics of computation and information distance
\newblock {\em Proc. 25th ACM Symp. Theory of Computation}.
\newblock ACM Press, 1993, 21-30.

\bibitem[Daley, 1973a]{Da73}
R.P. Daley, Minimal program complexity of sequences 
with restricted resources, {\em Inform. Contr.},
23(1973), 301-312.

\bibitem[Daley, 1973b]{Da73b}
R.P. Daley, An example of information and computation-resource
trade-off, {\em J. Assoc. Comput. Mach.}, 20:4(1973), 687-695.

\bibitem[Daley, 1975]{Da75}
R.P. Daley, Minimal-program complexity of pseudo-recursive
and pseudo-random sequences, {\em Math. Systems Theory},
9(1975), 83-94.

\bibitem[Deutsch, 1985]{De85}
D. Deutsch,
\newblock Quantum theory, the Church-Turing 
principle and the universal quantum computer.
\newblock {\em Proc. Royal Society London}.
\newblock Vol. A400(1985), 97-117.

\bibitem[Feynman, 1985]{Fe85}
R. Feynman.
\newblock Quantum mechanical computers.
\newblock {\em Foundations of Physics}, 16(1986), 507-531.
\newblock (Originally published in {\em Optics News}, February 1985.)

\bibitem[G\'acs, 1974]{Ga74}
P.~G\'acs.
\newblock On the symmetry of algorithmic information.
\newblock {\em Soviet Math. Dokl.}, 15:1477--1480, 1974.
\newblock Correction, Ibid., 15:1480, 1974.

\bibitem[Fredkin \& Toffoli, 1982]{FT82}
E. Fredkin and T. Toffoli.
\newblock Conservative logic.
\newblock {\em Int. J. Theoret. Phys.}, 21(1982),219-253.

\bibitem[Keyes, 1988]{Ke88}
R.W. Keyes, {\em IBM J. Res. Dev.}, 32(1988), 24-28.

\bibitem[Landauer, 1961]{La61}
R.~Landauer.
\newblock Irreversibility and heat generation in the computing process.
\newblock {\em IBM J. Res. Develop.}, 5:183--191, 1961.

\bibitem[Landauer, 1988]{La88}
R.~Landauer,
Dissipation and noise immunity in computation and communication,
{\em Nature}, 335(1988), 779-784.

\bibitem[Lecerf, 1963]{Le63}
Y. Lecerf,
 Machines de {T}uring r\'eversibles. 
{R}\'ecursive insolubilit\'e en $n \in
  {N}$ de l'\'equation $u=\theta^n$, o\`u 
$\theta$ est un ``isomorphisme de codes'',
{\em Comptes Rendus}, 257(1963), 2597-2600.

\bibitem[Levine and Sherman, 1990]{LeSh90}
R.Y. Levine and A.T. Sherman,
A note on Bennett's time-space trade-off for reversible
computation, {\em SIAM J. Comput.}, 19:4(1990), 673-677.

\bibitem[Li and Vit\'anyi, 1992]{LV}
M. Li and P.M.B. Vit\'anyi,
Theory of thermodynamics of computation,
{\em Preliminary Proc. Physics of Computation Workshop},
held in October 2-4, 1992, Dallas, Texas, (unpublished),
and as extended abstract with the same title
in the published record of that meeting, {\em IEEE Proc. Physics
and Computation Workshop}, IEEE Computer Society Press, 1992,
pp. 42-46.

\bibitem[Li and Vit\'anyi, 1993]{LiVi93}
M. Li and P.M.B. Vit\'anyi.
\newblock {\em An Introduction to Kolmogorov 
Complexity and Its Applications}.
\newblock Springer-Verlag, New York, 1993.

\bibitem[Merkle, 1993]{Me93}
R.C. Merkle, Reversible electronic logic using switches,
{\em Nanotechnology}, 4(1993), 21-40.

\bibitem[Proc. PhysComp, 1981, 1992, 1994]{PC}
\newblock {Proc. 1981 Physics and Computation Workshop}.
\newblock {\em Int. J. Theoret. Phys.}, 21(1982).
\newblock {\em Proc. IEEE 1992 Physics and Computation Workshop}.
\newblock IEEE Computer Society Press, 1992.
\newblock {\em Proc. IEEE 1994 Physics and Computation Workshop}.
\newblock IEEE Computer Society Press, 1994.

\bibitem[Shor, 1994]{Sh94}
Shor,~P.,
Algorithms for quantum computation: Discrete log and factoring,
{\em Proc. 35th IEEE Symposium on
Foundations of Computer Science}, 1994, 124-134.



\bibitem[Burks, 1966]{Bu}
J.~von Neumann.
\newblock {\em Theory of Self-Reproducing Automata}.
\newblock A.W. Burks, Ed., Univ. Illinois Press, Urbana, 1966.


\bibitem[Zurek, 1989]{Zu89a}
W.H. Zurek.
\newblock Thermodynamic cost of computation, algorithmic complexity and the
  information metric.
\newblock {\em Nature}, 341:119--124, 1989.

\end{thebibliography}

\end{document}